\documentclass[10pt]{article}
\usepackage{amsmath}
\usepackage{amsfonts}
\usepackage{graphicx}
\usepackage{authblk}
\usepackage{latexsym,amssymb,enumerate,ulem,pifont,calc}

\makeatletter

\@addtoreset{equation}{section}
\makeatother

\begin{document}

\begin{titlepage}
\vskip 2.0 cm

\begin{center}  {\huge{\bf Conjugation Matters}} \\\vskip 0.5 cm {\Large{\bf Bioctonionic Veronese Vectors and Cayley-Rosenfeld Planes}}

\vskip 2.5 cm
{\large{\bf Daniele Corradetti$^1$}, {\bf Alessio Marrani$^2$}, {\bf David Chester$^3$} and {\bf Ray Aschheim$^{3}$}}

\vskip 1.0 cm

$^1${\sl
Departamento de Matem\'{a}tica, Universidade do Algarve,\\ Campus de Gambelas, 8005-139 Faro, Portugal\\
\texttt{a55499@ualg.pt}}\\

\vskip 0.5 cm

$^2${\sl Instituto de F\'{i}sica Teorica, Dep.to de F\'{i}sica,\\Universidad de Murcia, Campus de Espinardo, E-30100, Spain\\
\texttt{jazzphyzz@gmail.com}}\\

\vskip 0.5 cm

$^3${\sl Quantum Gravity Research,\\Los Angeles, California, CA 90290, USA\\
\texttt{DavidC@QuantumGravityResearch.org}\\\texttt{Raymond@QuantumGravityResearch.org}}\\

\vskip 0.5
cm

 \end{center}

\vskip 2.0 cm

\begin{abstract}
Motivated by the recent interest in Lie algebraic and geometric structures
arising from tensor products of division algebras and their relevance to
high energy theoretical physics, we analyze generalized bioctonionic
projective and hyperbolic planes. After giving a Veronese representation of
the complexification of the Cayley plane $\mathbb{O}P_{\mathbb{C}}^{2}$, we
present a novel, explicit construction of the bioctonionic Cayley-Rosenfeld
plane $\left( \mathbb{C}\otimes \mathbb{O}\right) P^{2}$, again by
exploiting Veronese coordinates. We discuss the isometry groups of all
generalized bioctonionic planes, recovering all complex and real forms of
the exceptional groups $F_{4}$ and $E_{6}$, and characterizing such planes
as symmetric and Hermitian symmetric spaces. We conclude by discussing some
possible physical applications.
\end{abstract}
\vspace{24pt} \end{titlepage}

%{\parskip -6pt }

%\parskip 9pt
\newpage \tableofcontents \newpage

%%%%%%%%%%%%%%%%%%%%%%%%%%%%%%%%%%%%%%%%%%%%%%%%%%%%%%%%%%%%%%%%%%%%%%%%%%%%%%%%%%%%%%%%%%%%%%%%%%%%%%%%%%%%%%%%

\section{Introduction}

Recently, attention has been focused by theoretical physicists on Lie
structures arising from tensor products of division algebras, and on their
geometrical characterizations. It is well known that Tits-Freudenthal magic
square has numerous applications in super Yang-Mills and supergravity
theories \cite{GST,key-5,key-6}. At the same time, extensions of the
Standard Model based on non-division algebras resulting from the
Cayley-Dickson and their tensor product was proposed in a recent article by
Masi \cite{Masi G2}. On a different, but convergent, path, Todorov,
Dubois-Violette \cite{Todorov1} and Krasnov \cite{Krasnov} characterized the
Standard Model gauge group $G_{SM}$ as a subgroup of the automorphism group
of the exceptional Jordan algebra \textsc{$\mathfrak{J}$}$_{3}\left( \mathbb{%
O}\right) $, while Boyle \cite{Boyle, Corradetti} pointed to its
complexification $\mathfrak{J}_{3}^{\mathbb{C}}\left( \mathbb{O}\right) $.
It is well known that the group of determinant preserving automorphisms
(also named \textit{reduced structure group}) of the complexification of the
exceptional Jordan algebra is related to the collineations of the
complexification of the \textit{Cayley plane} $\mathbb{\mathbb{O}}P_{\mathbb{%
C}}^{2}$ (namely, the complexification of the octonionic projective plane)
and to $E_{6}\left( \mathbb{C}\right) $, while the group of trace preserving
automorphisms of the same Jordan algebra is related with the isometries of
the Cayley plane and to $F_{4}\left( \mathbb{C}\right) $. While $F_{4}$ does
not have complex representations, the exceptional Lie Group $E_{6}$ is
historically well known as a candidate for Grand Unification Theories \cite%
{Gursey:1975ki}.

Complex and real forms of $E_{6}$ are also notoriously related to symmetries
of the \textit{bioctonionic Rosenfeld plane} $\mathbb{\mathbb{\left( \mathbb{%
C}\otimes \mathbb{O}\right) }}P^{2}$. In a series of seminal papers resumed
and completed in \cite{Rosenfeld Group2,key-7}, Rosenfeld linked all the Lie
groups arising from Tits-Freudenthal magic square with the groups of
collineations and elliptic motions of generalised projective planes defined
over tensor product of division algebras and their split versions.
Consequently, he identified what are now known as \textit{Rosenfeld planes}
as symmetric spaces and Hermitian symmetric spaces over compact and non
compact form of real Lie groups. However, in his articles Rosenfeld left
some ambiguity in the definition of the octonionic part of the construction
: due to the lack of associativity of octonions, the usual identification of
the plane through the quotient of a module could not indeed be pursued;
moreover, when the resulting tensor algebra is not a composition algebra,
neither a direct completion of a generalised affine plane could be worked
out. A well known identification, due to Jordan, von Neumann and Wigner \cite%
{Jordan}, relates points of the projective plane $\mathbb{K}P^{2}$ with
rank-1 idempotent elements of the Jordan algebra $\mathfrak{J}_{3}\left(
\mathbb{K}\right) $ of Hermitian three by three matrices over the division
algebra $\mathbb{K}$. This provides a generalisation, but when defining a
Jordan algebra is not possible, the construction breaks down (see e.g. the
discussion at the end of Sec. 5 of \cite{magic-pyr}, and Refs.
therein).\medskip

This paper is devoted to highlight the crucial role played by conjugation
(and related norm) in determining the algebraic-geometric structure of
projective plane (in a generalized sense) over tensor products of division
algebras. In the present paper we will focus on the bioctonionic algebra $%
\mathbb{C}\otimes \mathbb{O}$, pointing out how its composition nature
strictly depends on the conjugation being considered. For the first time, we
will employ \textit{Veronese coordinates} over the bioctonions in order to
describe suitable real forms of the bioctonionic Rosenfeld plane; this will
pave the way to the treatment of more complicated generalized projective
spaces, which we plan to deal with in future works. We will present an
alternative, simple construction, based on the definition of \textit{%
Veronese vectors} over the bioctonions, that allows an explicit description
of two generalised projective planes over the algebra of bioctonions $%
\mathbb{C}\otimes \mathbb{O}$ that are of the most interest: the
complexification of the octonionic projective plane or\textit{\ Cayley plane}
$\mathbb{\mathbb{O}}P_{\mathbb{C}}^{2}$, and the \textit{bioctonionic
Rosenfeld plane} $\mathbb{\mathbb{\left( \mathbb{C}\otimes \mathbb{O}\right)
}}P^{2}$ . In Sec. \ref{2} we introduce the algebra of bioctonions with its
linear structure, conjugations and norms over both fields of real numbers $%
\mathbb{R}$ and complex numbers $\mathbb{C}$. In Sec. \ref{3} we explicitly
construct $\mathbb{\mathbb{O}}P_{\mathbb{C}}^{2}$ and $\mathbb{\mathbb{%
\left( \mathbb{C}\otimes \mathbb{O}\right) }}P^{2}$ making use of Veronese
coordinates, while in the former case the construction has already appeared
in the literature (see e.g. \cite{SBG11}), in the latter case the
construction by means of Veronese vectors is new, and it exhibits some
non-trivial features. Indeed, on one hand the complexification of the Cayley
plane $\mathbb{\mathbb{O}}P_{\mathbb{C}}^{2}$ is derived from a composition
algebra, it respects the Moufang identities and can be considered as a
completion of a generalised affine plane over the bioctonionic algebra. On
the other hand, since bioctonions are not a composition algebra with respect
to a suitably defined real norm, the bioctonionic Rosenfeld plane $\mathbb{%
\mathbb{\left( \mathbb{C}\otimes \mathbb{O}\right) }}P^{2}$ violates the
basic axioms of projective geometry, and it cannot be considered as an
extension and completion of a would-be affine Rosenfeld plane. Then, in Sec. %
\ref{4} we exploit the relation between Veronese vectors and simple, rank-3
Jordan algebras, and we thus identify $\mathbb{\mathbb{O}}P_{\mathbb{C}}^{2}$
with the space of rank-1 idempotent elements the complexification of the
exceptional Jordan Algebra $\mathfrak{J}_{3}^{\mathbb{C}}\left( \mathbb{O}%
\right) $. In Sec. \ref{5} we proceed to analyze the group of motions of the
generalised projective planes, recovering $F_{4}$ as the isometry group of
complexification of the Cayley plane $\mathbb{\mathbb{O}}P_{\mathbb{C}}^{2}$%
, and $E_{6}$ as the isometry group of the bioctonionic Rosenfeld plane $%
\mathbb{\mathbb{\left( \mathbb{C}\otimes \mathbb{O}\right) }}P^{2}$. Sec. %
\ref{6} then deals with a systematic definition of bioctonionic planes as
symmetric and Hermitian symmetric spaces, retrieving all the real forms of $%
F_{4}$ and $E_{6}$. Finally, in Sec. \ref{7} we discuss some possible
applications of $F_{4}$ and $E_{6}$ and the above related geometrical
structures to high energy theoretical physics. An outlook and comments on
further future developments are given in Sec. \ref{8}, which concludes the
paper.

\section{\label{2}The Algebra of Bioctonions}

\begin{figure}[tbp]
\centering{}\includegraphics[scale=0.07]{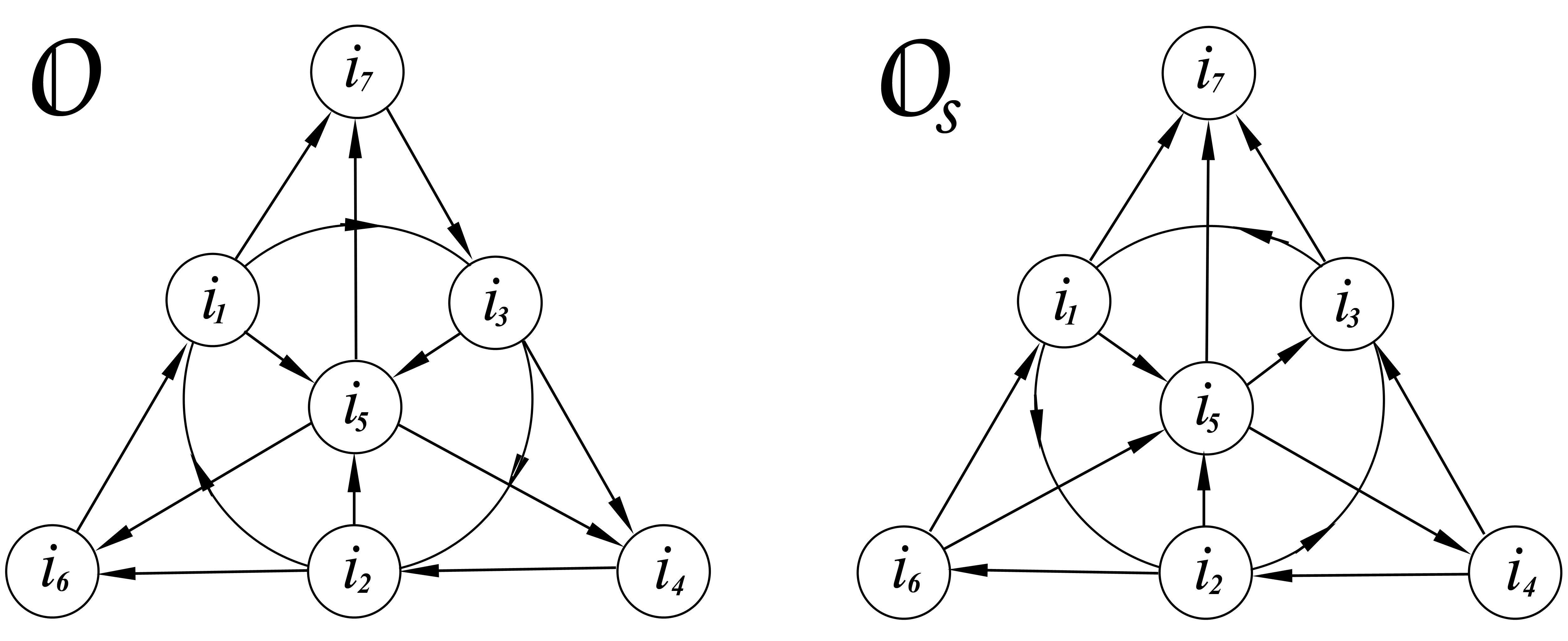}
\caption{Multiplication rule of octonions $\mathbb{O}$ (\textit{left}) and
of split-octonions $\mathbb{O}_{s}$ (\textit{right}) as real vector space $%
\mathbb{R}^{8}$ in the basis $\left\{ i_{0}=1,i_{1},...,i_{7}\right\} $. In
the case of the octonions $i_{0}^{2}=1$ and $i_{k}^{2}=-1$ for $k=1...7$,
while in the case of split-octonions $i_{k}^{2}=1$, for $k=1,2,3$ and $%
i_{k}^{2}=-1$ for $k\neq 1,2,3$. By exchanging the indices $2\leftrightarrow
3$ and $4\leftrightarrow 5$, the octonionic multiplication corresponds to
the Cayley-Graves' one \protect\cite{Cayley, Graves}, recently discussed in
\protect\cite{GKLY}.}
\label{fig:octonion fano plane-1}
\end{figure}

Let the octonions $\mathbb{O}$ be the only non-associative normed division
algebra with $\mathbb{O}_{s}$ as its split version, and let $\mathbb{C}$ be
the algebra of complex numbers and $\mathbb{C}_{s}$ its split algebra. We
then define the algebra of \textit{bioctonions} as the complexification of
the algebra of Octonions, i.e. as the tensor product $\mathbb{C}\otimes
\mathbb{O}$ or, equivalently, as $\mathbb{C}\otimes \mathbb{O}_{s}$. Since $%
\mathbb{O}$ is an alternative algebra and $\mathbb{C}$ is a commutative
algebra, then $\mathbb{C}\otimes \mathbb{O}$ is an \textit{alternative
algebra}. In the following sections we will work with the $\mathbb{R}^{16}$
and the $\mathbb{C}^{8}$ decomposition of bioctonions. In the $\mathbb{R}%
^{16}$ decomposition, an element of the bioctonionic algebra is given by
\begin{equation}
b:={\sum_{\alpha =0}^{7}}\left( x^{\alpha }+\text{i}y^{\alpha }\right)
i_{\alpha },  \label{eq:decomposition octonions}
\end{equation}%
where $x^{\alpha },y^{\alpha }\in \mathbb{R}$, the imaginary unit commutes
with the octonionic units , i.e. $\text{i}i_{\alpha }=i_{\alpha }\text{i}$,
the multiplication rules of $i_{\alpha }$ are given by the Fano plane in Fig.%
\ref{fig:octonion fano plane-1} left or right if bioctonions are considered
as $\mathbb{C}\otimes \mathbb{O}$ or as $\mathbb{C}\otimes \mathbb{O}_{s}$
respectively. Rewriting (\ref{eq:decomposition octonions}) we obtain the $%
\mathbb{C}^{8}$ decomposition
\begin{equation}
b:=\sum_{\alpha =0}^{7}z^{\alpha }i_{\alpha },
\label{eq:complex decomposition}
\end{equation}%
where $z^{\alpha }\in \mathbb{C}$. The two decompositions (\ref%
{eq:decomposition octonions}) and (\ref{eq:complex decomposition}) highlight
two different vector space structures available on the algebra of
bioctonions: the first is over the field of the real numbers $\mathbb{R}$
and is of dimension 16, while the second is over the complex field $\mathbb{C%
}$ and has complex dimension $8$. It is worth noting that the algebra of
bioctonions is not a division algebra\footnote{%
The treatment of the Veronese vectors over algebras containing zero-divisors
has been given e.g. in \cite{Chaput}, in which a suitably generalized
Veronese map is proposed (see Th. 5.2 therein).}, e.g. $\left( \text{i}%
i_{\alpha }+1\right) \left( \text{i}i_{\alpha }-1\right) =0$.

\subsection{Complex Norm}

Considering the bioctonions $\mathbb{C}\otimes \mathbb{O}$ as a complex
vector space, it is natural to define a complex norm. Let $b=z\otimes w$ a
bioctonion with $z\in \mathbb{C}$ and $w\in \mathbb{O}$, then its\textit{\
octonionic conjugate} $b^{\ast }$ is the element $b^{\ast }=z\otimes w^{\ast
}$, where $w^{\ast }\in \mathbb{O}$ is the conjugate of $\mathbb{O}$.
Applying the complex decomposition (\ref{eq:decomposition octonions}), then
the octonionic conjugate of $b$ has the form
\begin{equation}
b^{\ast }:=z^{0}-{\sum_{\alpha =1}^{7}}z^{\alpha }i_{\alpha }.
\end{equation}%
We then define an \textit{octonionic inner product} $\left\langle \cdot
,\cdot \right\rangle _{\mathbb{O}}$ over $\mathbb{C}\otimes \mathbb{O}$ as
\begin{equation}
\left\langle b_{1},b_{2}\right\rangle _{\mathbb{O}%
}:=z_{1}^{0}z_{2}^{0}+...+z_{1}^{7}z_{2}^{7}\in \mathbb{C},
\end{equation}%
where $z_{1}^{\alpha },z_{2}^{\alpha }\in \mathbb{C}$ are the complex
coefficients of $b_{1}$ and $b_{2}$ respectively. The octonionic inner
product induces a \textit{complex norm} $N\left( \cdot \right) $ in $\mathbb{%
C}$, given as
\begin{equation}
N\left( b\right) :=\left\langle b,b\right\rangle _{\mathbb{O}}\in \mathbb{C},
\end{equation}%
i.e. $N\left( b\right) =\left( z^{0}\right) ^{2}+...+\left( z^{7}\right)
^{2}=bb^{\ast }=b^{\ast }b$. The complex norm $N$ is a non degenerate
quadratic form over the complex vector space $\mathbb{C}\otimes \mathbb{O}$.
Moreover, $b$ is a zero divisor if and only if $N\left( b\right) =0$. In
respect to the complex norm $N$ we also have $N\left( \lambda b\right)
=\lambda ^{2}N\left( b\right) $ for every $\lambda \in \mathbb{C}$, and
\begin{equation}
N\left( b_{1}b_{2}\right) =N\left( b_{1}\right) N\left( b_{2}\right) ,
\end{equation}%
and therefore $\mathbb{C}\otimes \mathbb{O}$ is a \textit{composition algebra%
} with respect to the complex norm $N$.\medskip

\textbf{Remark 1}. As the octonionic inner product induces an inner product
and a complex norm over $\mathbb{C}\otimes \mathbb{O}$, also its split
version gives rise to a split-octonionic inner product with a norm. Even
though $\mathbb{C}\otimes \mathbb{O}$ and $\mathbb{C}\otimes \mathbb{O}_{s}$
give rise to the same bioctonionic algebra, we will write $\mathbb{C}\otimes
\mathbb{O}_{s}$ when we will intend the bioctonionic algebra equipped with
the split octonionic inner product and its norm.

\subsection{Real Norm}

We also define a real norm given by the \textit{bioctonionic conjugation},
i.e. $\overline{b}^{\ast }=\overline{z}\otimes w^{\ast }$, where $w^{\ast }$
is the octonionic conjugate of $w$ in $\mathbb{O}$ and $\overline{z}$ is the
complex conjugate of $z$ in $\mathbb{C}$. Consequently the inner product $%
\left\langle \cdot ,\cdot \right\rangle _{\mathbb{C}\otimes \mathbb{O}}$ is
defined as

\begin{equation}
\left\langle b_{1},b_{2}\right\rangle _{\mathbb{C}\otimes \mathbb{O}}:=%
\overline{z}_{1}^{0}z_{2}^{0}+...+\overline{z}_{1}^{7}z_{2}^{7},
\end{equation}%
and induce a \textit{real norm} $\left\Vert \cdot \right\Vert $ that is the
sum of the norms of the complex coefficients of $b$, i.e.
\begin{equation}
\left\Vert b\right\Vert ^{2}:=\left\langle b,b\right\rangle _{\mathbb{C}%
\otimes \mathbb{O}}=\left\vert z^{1}\right\vert ^{2}+...+\left\vert
z^{7}\right\vert ^{2}\in \mathbb{R}.
\end{equation}%
Since $\left\Vert ab\right\Vert ^{2}\neq \left\Vert a\right\Vert
^{2}\left\Vert b\right\Vert ^{2}$ then $\mathbb{C}\otimes \mathbb{O}$ is not
a \textit{composition algebra}\emph{\ }in respect to the real norm.

\subsection{Automorphisms}

Since the automorphisms of $\mathbb{C}$ are isomorphic to\footnote{%
We discard the so-called `wild' automorphisms \cite{Automorph}.} $\mathbb{Z}%
_{2}$ \cite{Baez}, and the automorphisms of $\mathbb{O}$ are isomorphic to
the exceptional Lie group $G_{2}$, i.e. $\text{Aut}\left( \mathbb{O}\right)
\cong G_{2},$ the group of automorphisms of the algebra of bioctonions is
isomorphic to

\begin{equation}
\text{Aut}\left( \mathbb{C}\otimes \mathbb{O}\right) =\mathbb{Z}_{2}\times
G_{2},
\end{equation}%
and consequently the Lie algebra of derivations is isomorphic to that of
octonions, i.e. $\mathfrak{der}\left( \mathbb{C}\otimes \mathbb{O}\right)
\cong \mathfrak{g}_{2}$.

\section{\label{3}Veronese Vectors over Bioctonions}

In this section we define explicitly two bioctonionic planes making use of
Veronese coordinates. In the case of the projective and hyperbolic planes on
octonions, i.e. $\mathbb{O}P^{2}$ and $\mathbb{O}H^{2}$ respectively, the
construction is well known \cite{SBG11}. The projective plane over the
octonions with real coefficients $\mathbb{R}\otimes \mathbb{O}\equiv \mathbb{%
O}_{\mathbb{R}}$ has been studied extensively in \cite{Tits53, Freud54}. A
rigorous definition of the octonionic projective plane, and the proof that
its automorphism group is a simple group of type $E_{6}$ in all
characteristics, can be found in \cite{Chaput} (see Th. 5.1 therein).

In the present paper, we study the complexification of the Cayley Plane $OP_{%
\mathbb{C}}^{2}$ and the bioctonionic Rosenfeld plane $\left( \mathbb{C}%
\otimes \mathbb{O}\right) P^{2}$\textsc{.} In the former case, we make use
of the complex vector space structure of $\mathbb{C}\otimes \mathbb{O}$, and
therefore of the octonionic conjugation $b^{\ast }$, the octonionic inner
product $\left\langle \cdot ,\cdot \right\rangle _{\mathbb{O}}$ and complex
norm $N\left( \cdot \right) $. In the latter case, we rely on the real
vector space structure of $\mathbb{C}\otimes \mathbb{O}$, and therefore on
the bioctonionic conjugation $\overline{b}^{\ast }$, the bioctonionic inner
product $\left\langle \cdot ,\cdot \right\rangle _{\mathbb{\mathbb{C}\otimes
\mathbb{O}}}$ and real norm $\left\Vert \cdot \right\Vert $; we will provide
a Veronese representation of the bioctonionic Rosenfeld plane $\left(
\mathbb{C}\otimes \mathbb{O}\right) P^{2}$ even if the bioctonions are not a
composition algebra with respect to the real norm $\left\Vert \cdot
\right\Vert $

\subsection{The Complexification of the Cayley Plane $\mathbb{O}P_{\mathbb{C}%
}^{2}$}

Let $V\cong \mathbb{\left( \mathbb{C}\otimes \mathbb{O}\right) }^{3}\times
\mathbb{C}^{3}$ be a complex vector space, with elements $\omega $ of the
form
\begin{equation}
\left( b_{\nu };\lambda _{\nu }\right) _{\nu }=\left(
b_{1},b_{2},b_{3};\lambda _{1},\lambda _{2},\lambda _{3}\right) ,
\end{equation}%
where $b_{\nu }\in \mathbb{\mathbb{C}\otimes \mathbb{O}}$, $\lambda _{\nu
}\in \mathbb{C}$ for $\nu =1,2,3$. A vector $\omega \in V$ is called \textit{%
Veronese} iff

\begin{align}
\lambda _{1}b_{1}^{\ast }& =b_{2}b_{3},\,\,\lambda _{2}b_{2}^{\ast
}=b_{3}b_{1},\,\,\lambda _{3}b_{3}^{\ast }=b_{1}b_{2},
\label{eq:Veronese conditions1} \\
N\left( b_{1}\right) & =\lambda _{2}\lambda _{3},\,N\left( b_{2}\right)
=\lambda _{3}\lambda _{1},N\left( b_{1}\right) =\lambda _{1}\lambda _{2}.
\label{eq:Veronese conditions2}
\end{align}%
Let the set $H\subset V$ be the set of Veronese vectors inside $V$. Since $%
\mathbb{C}$ is commutative and $\lambda ^{\ast }=\lambda $, since $\lambda
b=b\lambda $ and $N\left( \lambda b\right) =\lambda ^{2}N\left( b\right) $
when $\lambda \in \mathbb{C}$, then if $\omega $ is a Veronese vector, all
complex multiples $\mathbb{C\omega }$ are again Veronese vectors, i.e. if $%
\omega \in H$ then $\mu \omega \in H$ when $\mu \in \mathbb{C}$. We then
define the \textit{complexified octonionic plane} or \textit{%
complexification of the Cayley plane} $\mathbb{O}P_{\mathbb{C}}^{2}$ as the
set of 1-dimensional complex subspaces $\mathbb{C}\omega $ that we will call
\textit{points} of the plane, i.e.
\begin{equation}
\mathbb{O}P_{\mathbb{C}}^{2}:=\left\{ \mathbb{C}\omega :\omega \in
H\smallsetminus \left\{ 0\right\} \right\} .  \label{compl}
\end{equation}

\subsubsection{Complexified Cayley Lines}

Lines in $\mathbb{O}P_{\mathbb{C}}^{2}$ are orthogonal subspaces of a point
of the plane. Therefore, let $\mathbb{C}\omega $ be a point in $\mathbb{O}P_{%
\mathbb{C}}^{2}$, we define the line $\ell $ as the orthogonal subspace
\begin{equation}
\ell :=\omega ^{\perp }=\left\{ \upsilon \in V:\beta \left( \upsilon ,\omega
\right) =0\right\} ,
\end{equation}%
where the complex bilinear form $\beta $ is given by

\begin{equation}
\beta \left( \upsilon ,\omega \right) :={\sum_{\nu =1}^{3}}\left(
\left\langle b_{\nu }^{1},b_{\nu }^{2}\right\rangle _{\mathbb{O}}+\lambda
_{\nu }^{1}\lambda _{\nu }^{2}\right) ,
\end{equation}%
with $\upsilon ,\omega \in V$, of coordinates $\left( b_{\nu }^{1};\lambda
_{\nu }^{1}\right) _{\nu }$,$\left( b_{\nu }^{2};\lambda _{\nu }^{2}\right)
_{\nu }$ respectively.

\subsubsection{Elliptic and Hyperbolic Polarity on $\mathbb{O}P_{\mathbb{C}%
}^{2}$}

Since every point $\mathbb{C}\omega $ of the plane defines an orthogonal
line $\omega ^{\perp }\subset \mathbb{O}P_{\mathbb{C}}^{2}$ and, as
converse, every line defines a point, we call \textit{standard elliptic
polarity} $\pi ^{+}$ the involutive map that corresponds points to lines and
lines to points through orthogonality, i.e.
\begin{equation}
\pi ^{+}\left( \omega \right) :=\omega ^{\perp },\pi ^{+}\left( \omega
^{\perp }\right) :=\omega ,
\end{equation}%
using the complex bilinear form $\beta \left( \cdot ,\cdot \right) $ so that
\begin{equation}
\omega \longrightarrow \left\{ \beta \left( \cdot ,\omega \right) =0\right\}
.
\end{equation}%
Explicitly, $\beta \left( \upsilon ,\omega \right) =0$ when
\begin{equation}
b_{1}^{1}b_{1}^{2\ast }+b_{2}^{1}b_{2}^{2\ast }+b_{3}^{1}b_{3}^{2\ast
}+\lambda _{1}^{1}\lambda _{1}^{2}+\lambda _{2}^{1}\lambda _{2}^{2}+\lambda
_{3}^{1}\lambda _{3}^{2}=0,
\end{equation}%
where, as before, we intended, $\left( b_{\nu }^{1};\lambda _{\nu
}^{1}\right) _{\nu }$ and $\left( b_{\nu }^{2};\lambda _{\nu }^{2}\right)
_{\nu }$ as the coordinates of $\upsilon ,\omega \in V$. We also define an
\textit{hyperbolic polarity }$\pi ^{-}$\textit{\ }as the involutive map
between points and lines which still has
\begin{equation}
\pi ^{-}\left( \omega \right) :=\omega ^{\perp },\pi ^{-}\left( \omega
^{\perp }\right) :=\omega ,
\end{equation}%
but through the use of the bilinear form $\beta _{-}$ which has a change of
sign in the last coordinate, i.e. $\beta _{-}\left( \upsilon ,\omega \right)
=0$ when
\begin{equation}
b_{1}^{1}b_{1}^{2\ast }+b_{2}^{1}b_{2}^{2\ast }-b_{3}^{1}b_{3}^{2\ast
}+\lambda _{1}^{1}\lambda _{1}^{2}+\lambda _{2}^{1}\lambda _{2}^{2}-\lambda
_{3}^{1}\lambda _{3}^{2}=0.
\end{equation}%
The projective plane equipped with the bilinear form $\beta _{-}$ and the
hyperbolic polarity\emph{\ }$\pi ^{-}$ it will be called the \textit{%
complexified hyperbolic Cayley plane} $\mathbb{O}H_{\mathbb{C}}^{2}$.

\subsubsection{Complexified Octonionic Affine Plane}

\begin{figure}[tbp]
\centering{}\includegraphics[scale=0.4]{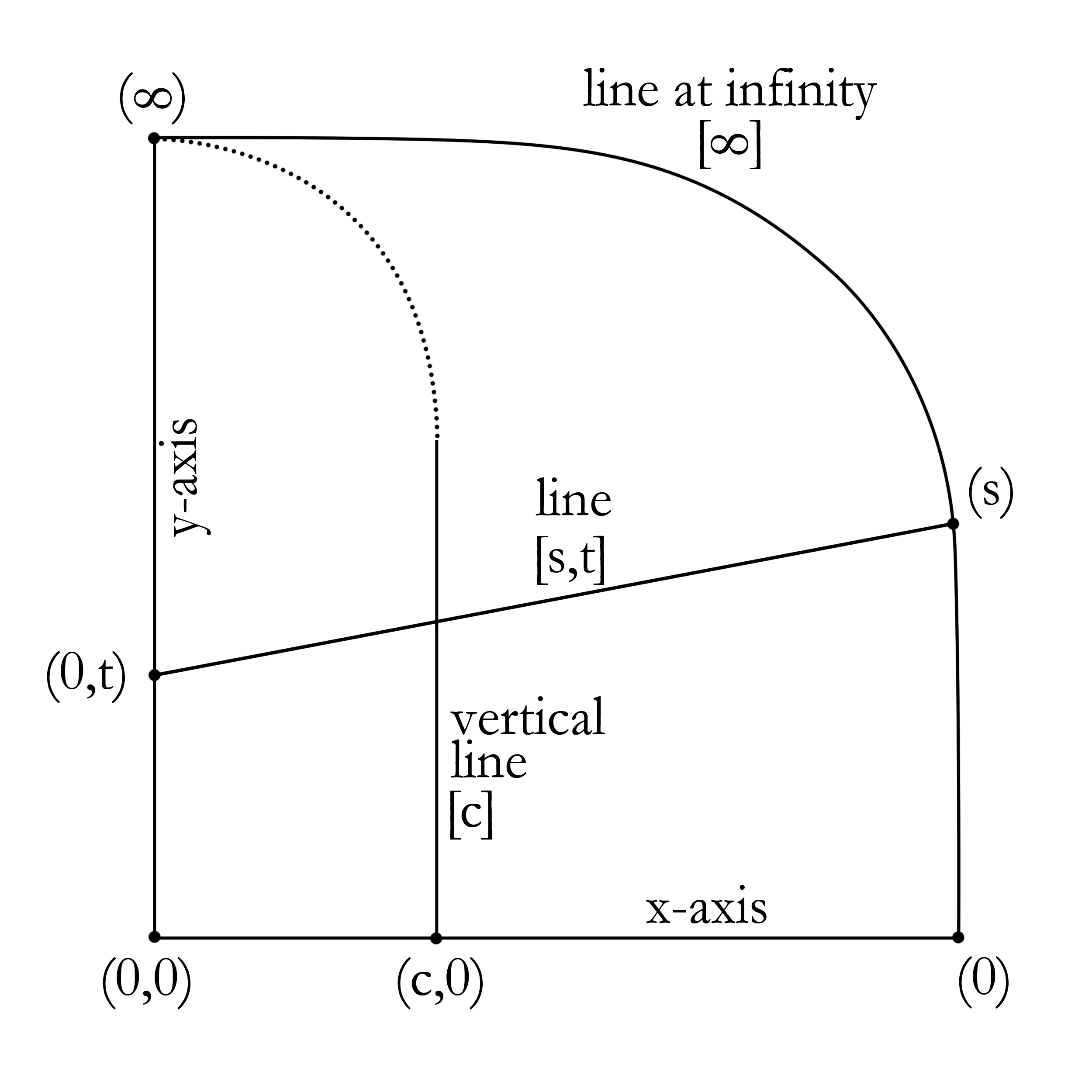}
\caption{Representation of the affine plane: $\left(0,0\right)$ represents
the origin, $\left(0\right)$ the point at the infinity on the $x$-axis, $%
\left(s\right)$ is the point at infinity of the line $\left[s,t\right]$ of
slope $s$ while $\left(\infty\right)$ is the point at the infinity on the $y$%
-axis and of vertical lines $\left[c\right]$.}
\label{fig:affine-plane}
\end{figure}

In analogy to the classic case, the complexification of the Cayley plane can
also be seen as the completion and topological compactification of a \textit{%
bioctonionic affine plane} $\left( \mathbb{C}\otimes \mathbb{O}\right) A^{2}$%
, but, since $\mathbb{C}\otimes \mathbb{O}$ is not a division algebra, the
strict set of axioms and results of affine geometry are not valid on this
plane. Indeed, the map from $\left( \mathbb{C}\otimes \mathbb{O}\right) ^{2}$
to $V$ defined as
\begin{equation}
\left( x,y\right) \mapsto \mathbb{C}\left( x,y^{\ast },yx^{\ast };N\left(
y\right) ,N\left( x\right) ,1\right) ,  \label{eq:map affine-projective}
\end{equation}%
sends elements $\left( x,y\right) \in \left( \mathbb{C}\otimes \mathbb{O}%
\right) ^{2}$ into Veronese vectors, and therefore to $\mathbb{O}P_{\mathbb{C%
}}^{2}$ establishing a correspondence between points of the complexification
of the Cayley plane and elements of an affine plane $\left( \mathbb{C}%
\otimes \mathbb{O}\right) A^{2}$ with bioctonionic coordinates $\left(
x,y\right) $. To show that $\mathbb{O}P_{\mathbb{C}}^{2}$ is a completion of
the affine plane we add two sets of point of coordinates $\left( x\right) $
and $\left( \infty \right) $ that will extend the map to cover the whole $%
\mathbb{O}P_{\mathbb{C}}^{2}$ as follows :

\begin{align}
\left( x\right) & \mapsto \mathbb{C}\left( 0,0,x;N\left( x\right)
,1,0\right) , \\
\left( \infty \right) & \mapsto \left( 0,0,0;1,0,0\right) ,
\end{align}%
where $x\in \mathbb{C}\otimes \mathbb{O}$. Finally, in order to give a
complete picture, a line in the affine plane $\left( \mathbb{C}\otimes
\mathbb{O}\right) A^{2}$ will be given by
\begin{equation}
\left[ s,t\right] :=\left\{ \left( x,sx+t\right) :x,t\in \mathbb{\mathbb{C}%
\otimes \mathbb{O}}\right\} ,
\end{equation}%
where $s$ $\in \mathbb{C}\otimes \mathbb{O}$ is the \textit{slope} of the
line. Vertical lines will be indicated as $\left[ c\right] :=\left\{
c\right\} \times \left( \mathbb{C}\otimes \mathbb{O}\right) $. As for the
previous case, lines of the affine plane $\left( \mathbb{C}\otimes \mathbb{O}%
\right) A^{2}$ have a correspondence with lines of the projective plane $%
\mathbb{O}P_{\mathbb{C}}^{2}$ through the map

\begin{align}
\left[ s,t\right] & \mapsto \mathbb{C}\left( s^{\ast }t,-t^{\ast
},-s;1,N\left( s\right) ,N\left( t\right) \right) ^{\perp }, \\
\left[ c\right] & \mapsto \mathbb{C}\left( -c,0,0;0,1,N\left( c\right)
\right) ^{\perp },
\end{align}%
where $s,t,c\in \mathbb{C}\otimes \mathbb{O}$. The correspondence is
bijective if we add a line $\left[ \infty \right] $ that will correspond to
\begin{equation}
\left[ \infty \right] \mapsto \mathbb{C}\left( 0,0,0;0,0,1\right) ^{\perp }.
\end{equation}%
\medskip

\textbf{Remark 2}. The complexification of the octonionic affine plane does
not satisfy the axioms of usual affine geometry. It is critical the
existence of \textit{adjacent points}, i.e. points that are separated by a
singular affine vector $\left( v_{1},v_{2}\right) $ such that $%
(av_{1},av_{2})=0$, with $a,v_{1},v_{2}\in \mathbb{C}\otimes \mathbb{O}$.
Between two adjacent points passes more than one line, that are called
\textit{adjacent} themselves. Two lines that are not adjacent but that can
be transformed into adjacent through a translation are called \textit{%
diverging lines}. So any two given lines in this plane might be \textit{%
incident}, \textit{parallel}, \textit{coincident}, \textit{adjacent}, or
\textit{divergent} (cfr. \cite{Rosenfeld Group2}).\medskip

\textbf{Remark 3}. The counting of (complex) dimensions goes as follows. By
construction, dim$_{\mathbb{C}}V=27$. Since $\mathbb{C}\otimes \mathbb{O}$
is composition with respect to the complex norm $N\left( \cdot \right) $,
only 10 relations are independent out of all 27 ones (\ref{eq:Veronese
conditions1})-(\ref{eq:Veronese conditions2}) defining Veronese vectors;
thus dim$_{\mathbb{C}}H$ $=17$. Finally, from (\ref{compl}) dim$_{\mathbb{C}%
}(\mathbb{O}P_{\mathbb{C}}^{2})=$dim$_{\mathbb{C}}H-1=16$, as expected.

\subsection{The Bioctonionic Rosenfeld Plane $\left(\mathbb{C}\otimes\mathbb{%
O}\right)P^{2}$}

We now replicate the same construction, considering as a starting point the
real vector space $V\cong \mathbb{\left( \mathbb{C}\otimes \mathbb{O}\right)
}^{3}\times \mathbb{R}^{3}$ and the real norm over the bioctonionic algebra.
In this case elements of $V$ are of the form
\begin{equation}
\left( b_{\nu };\lambda _{\nu }\right) _{\nu }=\left(
b_{1},b_{2},b_{3};\lambda _{1},\lambda _{2},\lambda _{3}\right) ,
\end{equation}%
where $b_{\nu }\in \mathbb{\mathbb{C}\otimes \mathbb{O}}$, $\lambda _{\nu
}\in \mathbb{R}$ and $\nu =1,2,3$. A vector $\omega \in V$ is called \textit{%
Veronese} iff

\begin{align}
\lambda _{1}\overline{b}_{1}^{\ast }& =b_{2}b_{3},\,\,\lambda _{2}\overline{b%
}_{2}^{\ast }=b_{3}b_{1},\,\,\lambda _{3}\overline{b}_{3}^{\ast }=b_{1}b_{2},
\label{eq:Veronese* conditions1} \\
\left\Vert b_{1}\right\Vert ^{2}& =\lambda _{2}\lambda _{3},\,\left\Vert
b_{2}\right\Vert ^{2}=\lambda _{3}\lambda _{1},\left\Vert b_{3}\right\Vert
^{2}=\lambda _{1}\lambda _{2}.  \label{eq:Veronese* conditions2}
\end{align}%
Let $H\subset V$ be the set of Veronese vectors. Since $\mathbb{R}$ is a
commutative field, $\mu ^{\ast }=\mu $ and $\left\Vert \mu b\right\Vert
^{2}=\mu ^{2}\left\Vert b\right\Vert ^{2}$ when $\mu \in \mathbb{R}$, given
a Veronese vector $\omega \in H$, all real multiples $\mathbb{R\omega }$ are
again Veronese vectors, i.e. $\mathbb{R}\omega \subset H$. We define the
\textit{bioctonionic Rosenfeld plane} $\mathbb{\left( \mathbb{C}\otimes
O\right) }P^{2}$ as the set of 1-dimensional subspaces $\mathbb{R}\omega $
that we will call \textit{points} of the plane, i.e.
\begin{equation}
\mathbb{\left( \mathbb{C}\otimes O\right) }P^{2}:=\left\{ \mathbb{R}\omega
:\omega \in H\smallsetminus \left\{ 0\right\} \right\} .
\end{equation}

\subsubsection{Rosenfeld Lines}

As in the case of the complexification of the Cayley plane, we define the
lines of the bioctonionic Rosenfeld plane $\mathbb{\left( \mathbb{C}\otimes
O\right) }P^{2}$ as orthogonal subspaces of a point through the extension of
the bioctonionic inner product. Therefore, let $\mathbb{R}\omega $ be a
point in $\mathbb{\left( \mathbb{C}\otimes \mathbb{O}\right) }P^{2}$, then
the line $\ell $ is the orthogonal subspace
\begin{equation}
\ell :=\omega ^{\perp }=\left\{ \upsilon \in V:\beta _{\mathbb{C}\otimes
\mathbb{O}}\left( \upsilon ,\omega \right) =0\right\} ,
\end{equation}%
where the bilinear form is defined as

\begin{equation}
\beta _{\mathbb{C}\otimes \mathbb{O}}\left( \upsilon ,\omega \right) :={%
\sum_{\nu =1}^{3}}\left( \left\langle b_{\nu }^{1},b_{\nu }^{2}\right\rangle
_{\mathbb{\mathbb{C}\otimes O}}+\lambda _{\nu }^{1}\lambda _{\nu
}^{2}\right) ,
\end{equation}%
with $\upsilon ,\omega \in V$ of coordinates $\left( b_{\nu }^{1};\lambda
_{\nu }^{1}\right) _{\nu }$,$\left( b_{\nu }^{2};\lambda _{\nu }^{2}\right)
_{\nu }$ respectively.

\subsubsection{Elliptic and Hyperbolic Polarity on $\mathbb{\left( \mathbb{C}%
\otimes \mathbb{O}\right) }P^{2}$}

Since every point $\mathbb{R}\omega $ defines an orthogonal line $\omega
^{\perp }\subset \mathbb{\left( \mathbb{C}\otimes \mathbb{O}\right) }P^{2}$
and every line defines a point, we call standard elliptic polarity $\pi ^{+}$
the involutive map that correspond points to lines and lines to points
through orthogonality, i.e.
\begin{equation}
\pi ^{+}\left( \omega \right) :=\omega ^{\perp },\pi ^{+}\left( \omega
^{\perp }\right) :=\omega ,
\end{equation}%
making use of the bilinear form $\beta _{\mathbb{C}\otimes \mathbb{O}}\left(
\cdot ,\cdot \right) $. Explicitly, $\beta _{\mathbb{C}\otimes \mathbb{O}%
}\left( \upsilon ,\omega \right) =0$ when
\begin{equation}
b_{1}^{1}\overline{b}_{1}^{2\ast }+b_{2}^{1}\overline{b}_{2}^{2\ast
}+b_{3}^{1}\overline{b}_{3}^{2\ast }+\lambda _{1}^{1}\lambda
_{1}^{2}+\lambda _{2}^{1}\lambda _{2}^{2}+\lambda _{3}^{1}\lambda _{3}^{2}=0,
\end{equation}%
where, as before, we intended, $\left( b_{\nu }^{1};\lambda _{\nu
}^{1}\right) _{\nu }$ and $\left( b_{\nu }^{2};\lambda _{\nu }^{2}\right)
_{\nu }$ as the coordinates of $\upsilon ,\omega \in V$.

As in the previous case, the elliptic polarity is not the only possible one,
indeed we define the \textit{hyperbolic polarity }$\pi ^{-}$\emph{\ }as the
involutive map between points and lines which still has
\begin{equation}
\pi ^{-}\left( \omega \right) :=\omega ^{\perp },\pi ^{-}\left( \omega
^{\perp }\right) :=\omega ,
\end{equation}%
but that uses the bilinear form $\beta _{\mathbb{C}\otimes \mathbb{O}}^{-}$
which has a change of sign in the last coordinate, i.e. $\beta _{\mathbb{C}%
\otimes \mathbb{O}}^{-}\left( \upsilon ,\omega \right) =0$ when
\begin{equation}
b_{1}^{1}\overline{b}_{1}^{2\ast }+b_{2}^{1}\overline{b}_{2}^{2\ast
}-b_{3}^{1}\overline{b}_{3}^{2\ast }+\lambda _{1}^{1}\lambda
_{1}^{2}+\lambda _{2}^{1}\lambda _{2}^{2}-\lambda _{3}^{1}\lambda _{3}^{2}=0.
\end{equation}%
The projective plane equipped with the bilinear form $\beta _{\mathbb{C}%
\otimes \mathbb{O}}^{-}$ instead of $\beta _{\mathbb{C}\otimes \mathbb{O}}$
it will be called the bioctonionic Rosenfeld hyperbolic plane $\mathbb{%
\left( \mathbb{C}\otimes \mathbb{O}\right) }H^{2}$.\medskip

\textbf{Remark 4}. Since $\mathbb{\mathbb{C}\otimes \mathbb{O}}$ is not a
composition algebra with respect of the real norm $\left\Vert \cdot
\right\Vert $, a map as in (\ref{eq:map affine-projective}) would not be
well defined. Therefore, we cannot consider the Bioctonionic Rosenfeld plane
as an extension and completion of an affine Rosenfeld plane.\medskip

\textbf{Remark 5}. The counting of (real) dimensions goes as follows. Since $%
\mathbb{\mathbb{C}\otimes \mathbb{O}}$ is not a composition algebra with
respect to the real norm $\left\Vert \cdot \right\Vert $, one must consider (%
\ref{eq:Veronese* conditions2}) and, say, the first of (\ref{eq:Veronese*
conditions1}), as independent relations out of the relations (\ref%
{eq:Veronese* conditions1})-(\ref{eq:Veronese* conditions2}) defining
Veronese vectors. This corresponds to $8\times 2$ $+1+1+1=19$ real
conditions out of the $8\times 2\times 3+1+1+1=51$ real relations (\ref%
{eq:Veronese* conditions1})-(\ref{eq:Veronese* conditions2}). Thus, the real
dimension of $\mathbb{\left( \mathbb{C}\otimes \mathbb{O}\right) }P^{2}$ is
given by dim$_{\mathbb{R}}\left( \mathbb{\left( \mathbb{C}\otimes \mathbb{O}%
\right) }P^{2}\right) =51-19=32$, as expected.

\section{\label{4}Veronese Vectors and Jordan Algebras}

A well known identification relates rank-1 idempotent elements of the Jordan
algebra $\mathfrak{J}_{3}\left( \mathbb{O}\right) $ with points of the
octonionic projective plane $\mathbb{O}P^{2}$. While this identification
still stands for the complexification of the Cayley plane $\mathbb{O}P_{%
\mathbb{C}}^{2}$, when it is applied to the Rosenfeld plane $\mathbb{\left(
\mathbb{C}\otimes \mathbb{O}\right) }P^{2}$ one obtains that $\mathfrak{J}%
_{3}\left( \mathbb{C}\otimes \mathbb{O}\right) $ is a simple Jordan algebra,
but not a formally real one (cfr. e.g. \cite{Baez}).

As we discussed above, Veronese vectors, defined by conditions in (\ref%
{eq:Veronese conditions1}), are an alternative and useful way to
characterize rank-1 idempotent elements of a Jordan algebra $\mathfrak{J}%
_{3}\left( \mathbb{K}\right) $, with $\mathbb{K}$ being any tensor product
of division algebras.

\subsection{$\mathbb{O}P_{\mathbb{C}}^{2}$}

In order to show such relation within $\mathbb{O}P_{\mathbb{C}}^{2}$, let $%
\omega $ be an element of $V\cong \mathbb{\left( \mathbb{C}\otimes \mathbb{O}%
\right) }^{3}\times \mathbb{C}^{3}$ with coordinates $\left(
b_{1},b_{2},b_{3};\lambda _{1},\lambda _{2},\lambda _{3}\right) $ and define
$A_{\omega }$ as the three by three bioctonionic matrix given by
\begin{equation}
V\ni \omega \longrightarrow A_{\omega }:=\left(
\begin{array}{ccc}
\lambda _{1} & b_{3} & b_{2}^{\ast } \\
b_{3}^{\ast } & \lambda _{2} & b_{1} \\
b_{2} & b_{1}^{\ast } & \lambda _{3}%
\end{array}%
\right) \in \mathbb{C}\otimes \mathfrak{J}_{3}\left( \mathbb{O}\right) .
\end{equation}%
Note that since the scalar field $\mathbb{C}$ commutes with the coefficient $%
b$, then all $\mathbb{C}$-multiples of the vector $\omega $ are sent in
multiple of the matrix $A_{\omega }$. Therefore the map is well defined and
induces a bijective map between points $\mathbb{C}\omega \in \mathbb{O}P_{%
\mathbb{C}}^{2}$ and subspaces of the form $\mathbb{C}A_{\omega }$. It is
here worth recalling that the \textit{cubic norm} $\mathcal{N}$ of a
non-zero element $A_{\omega }\in \mathbb{C}\otimes \mathfrak{J}_{3}\left(
\mathbb{O}\right) $ is defined in terms of the generalization of the
determinant for three by three matrices with not necessarily associative
elements\footnote{%
See e.g. the example 5 of \cite{Krut}, which actually is a simplified
version of the reduced cubic factor example in Sec. I.3.9 of \cite{McCrimmon}%
.} :%
\begin{equation}
\mathcal{N}\left( A_{\omega }\right) \equiv \text{det}\left( A_{\omega
}\right) :=\lambda _{1}\lambda _{2}\lambda _{3}-\lambda _{1}N(b_{1})-\lambda
_{2}N\left( b_{2}\right) -\lambda _{3}N\left( b_{3}\right) +2\text{Re}\left(
\left( b_{1}b_{2}\right) b_{3}\right) ,  \label{N}
\end{equation}%
where Re$\left( \left( b_{1}b_{2}\right) b_{3}\right) :=\frac{1}{2}\left(
\left( b_{1}b_{2}\right) b_{3}+\overline{b_{3}}\left( \overline{b_{2}}%
\overline{b_{1}}\right) \right) $, implying that det$\left( aA_{\omega
}\right) =a^{3}$det$\left( A_{\omega }\right) $, $\forall a\in \mathbb{C}$.
It should be remarked that the determinant is actually well defined, as one
can realize by recalling the \textit{Hamilton Cayley identity} (see e.g.
\cite{Yokota}), i.e.
\begin{equation}
A_{\omega }\circ A_{\omega }^{2}-\text{tr}\left( A_{\omega }\right)
A_{\omega }^{2}+\frac{1}{2}\left( \text{tr}\left( A_{\omega }\right) ^{2}-%
\text{tr}\left( A_{\omega }^{2}\right) \right) A_{\omega }=\text{det}\left(
A_{\omega }\right) I,
\end{equation}%
where $I$ is the three by three identity matrix, and $\circ $ is the Jordan
product $A\circ B:=\frac{1}{2}\left( AB+BA\right) $.

By further specifying that $\omega $ is a Veronese vector in $V$, namely
that $\omega \in H\subset V$, and thus by plugging the Veronese conditions (%
\ref{eq:Veronese conditions1})-(\ref{eq:Veronese conditions2}) into (\ref{N}%
), one obtains that the norm of the corresponding element $A_{\omega }$
vanishes :
\begin{equation}
\omega \in H\Rightarrow \mathcal{N}\left( A_{\omega }\right) =0.
\end{equation}%
Moreover, one can consider the image $A_{\omega }^{\sharp }$ of a non-zero
element $A_{\omega }$, with $\omega \in H$, under the so-called \textit{%
adjoint} ($\sharp $-)\textit{map} of $\mathbb{C}\otimes \mathfrak{J}%
_{3}\left( \mathbb{O}\right) $, which is given by (again, cf. e.g. example 5
of \cite{Krut})%
\begin{equation}
A_{\omega }^{\sharp }:=\left(
\begin{array}{ccc}
\lambda _{2}\lambda _{3}-N(b_{1}) & \overline{b_{2}}\overline{b_{1}}-\lambda
_{3}b_{3} & b_{3}b_{1}-\lambda _{2}\overline{b_{2}} \\
b_{1}b_{2}-\lambda _{3}\overline{b_{3}} & \lambda _{1}\lambda _{3}-N(b_{2})
& \overline{b_{3}}\overline{b_{2}}-\lambda _{1}b_{1} \\
\overline{b_{1}}\overline{b_{3}}-\lambda _{2}b_{2} & b_{2}b_{3}-\lambda _{1}%
\overline{b_{1}} & \lambda _{1}\lambda _{2}-N(b_{3})%
\end{array}%
\right) ,
\end{equation}%
and which, when recalling (\ref{eq:Veronese conditions1})-(\ref{eq:Veronese
conditions2}), can be realized to vanish (again!), thus implying that%
\begin{equation}
\omega \in H\Leftrightarrow A_{\omega }^{\sharp }=0.
\end{equation}%
Thus, by the Definition 11 of \cite{Krut} (namely, from the invariant
definition of the rank of an element of $\mathfrak{J}_{3}^{\mathbb{O}}$ \cite%
{Jac}), one obtains that the Veronese conditions (\ref{eq:Veronese
conditions1})-(\ref{eq:Veronese conditions2}) are an equivalent
characterization of the rank-1 elements of the complexification of the
exceptional Jordan algebra $\mathbb{C}\otimes \mathfrak{J}_{3}\left( \mathbb{%
O}\right) \equiv \mathfrak{J}_{3}^{\mathbb{C}}\left( \mathbb{O}\right) $
(see e.g. \cite{Jacobson} for an extensive analysis).\medskip

\textbf{Remark 6}. The definition (\ref{compl}) characterizes the points of $%
\mathbb{O}P_{\mathbb{C}}^{2}$ as complex `Veronese rays', thus obtaining a $%
16_{\mathbb{C}}$-dimensional subspace of the unique orbit of rank-1 elements
of $\mathfrak{J}_{3}^{\mathbb{C}}\left( \mathbb{O}\right) $. Any well
defined representative of such a 16-dimensional subspace has a fixed trace.

\subsection{$\mathbb{\left( \mathbb{C}\otimes \mathbb{O}\right) P}^{2}$}

The same argument, with a different ending point, applies to the
bioctonionic Rosenfeld plane. Given a Veronese vector $\omega $ in $V\cong
\mathbb{\left( \mathbb{C}\otimes \mathbb{O}\right) }^{3}\times \mathbb{R}%
^{3} $ with coordinates $\left( b_{1},b_{2},b_{3};\lambda _{1},\lambda
_{2},\lambda _{3}\right) $ we define $A_{\omega }$ as the three by three
bioctonionic Hermitian matrix as
\begin{equation}
\omega \longrightarrow A_{\omega }:=\left(
\begin{array}{ccc}
\lambda _{1} & b_{3} & \overline{b}_{2}^{\ast } \\
\overline{b}_{3}^{\ast } & \lambda _{2} & b_{1} \\
b_{2} & \overline{b}_{1}^{\ast } & \lambda _{3}%
\end{array}%
\right) .
\end{equation}%
Since the scalar field $\mathbb{R}$ commutes with the coefficient $b$, then
all $\mathbb{R}$-multiples of the vector $\omega $ are sent in multiple of
the matrix $A_{\omega }$. Therefore, the map is well defined and induces a
injective map between points $\mathbb{\mathbb{R}}\omega \in \mathbb{\left(
\mathbb{C}\otimes \mathbb{O}\right) P}^{2}$ and subspaces $\mathbb{\mathbb{R}%
}A_{\omega }\subset H_{3}\left( \mathbb{C}\otimes \mathbb{O}\right) $, the
algebra of $3\times 3$ matrices with $\left( \mathbb{C}\otimes \mathbb{O}%
\right) $-valued entries and Hermitian with respect to the bi-octonionic
conjugation; this time, the bioctonionic conjugation still allows $A_{\omega
}$ to be endowed with the structure of a simple Jordan algebra, but not of a
formally real one.

\section{\label{5}Real Forms of $F_{4}$ and $E_{6}$}

Symmetries of the generalised projective planes over the bioctonionic
algebra lead naturally to all the complex and real forms of the exceptional
group $E_{6}$ and $F_{4}$. To do so we will look to generalised
collineations, i.e. automorphisms of generalised planes that sends lines
into lines, and elliptic and hyperbolic motions, i.e. collineations over the
projective (hyperbolic) plane that preserve the elliptic (hyperbolic)
polarity. Sometimes, due to correspondence between idempotent Jordan
matrices and points in the projective plane, the collineation group of the
octonionic plane $\text{Coll}\left( \mathbb{O}P^{2}\right) $ is called $%
SL\left( 3,\mathbb{O}\right) $, while the elliptic polarity preserving group
$\text{Iso}\left( \mathbb{O}P^{2}\right) $ is identified with $SU\left( 3,%
\mathbb{O}\right) $ \cite{Dray}.

To recover the collineations group and the polarity preserving group of the
octonionic projective and hyperbolic space, one might proceed in a geometric
\cite{SBG11}, group algebraic \cite{Jacobson} and Lie algebraic way \cite%
{key-7}. We will represent the last one following Rosenberg focusing on the
Lie algebra of the collineation group $\text{Coll}\left( \mathbb{O}%
P^{2}\right) $ that is given by the direct sum the Lie Algebra given by the
group of automorphisms of the field, in this case $\text{Aut}\left( \mathbb{O%
}\right) =G_{2}$, and the algebra $\mathfrak{a}_{3}\left( \mathbb{O}\right) $
of three by three matrices on $\mathbb{O}$ and null trace, i.e. $\text{tr}%
\left( A\right) =0$. We therefore have
\begin{equation}
\mathfrak{coll}\left( \mathbb{O}P^{2}\right) =\mathfrak{g}_{2}\oplus
\mathfrak{a}_{3}\left( \mathbb{O}\right) .
\end{equation}%
A simple count on the dimension on the generators of the $\mathfrak{a}%
_{3}\left( \mathbb{O}\right) $ algebra, imposing the null trace condition,
leads $8$ entries of dimension $8$ and therefore $\text{dim}{}_{\mathbb{R}}%
\mathfrak{a}_{3}=64$, that brings to
\begin{equation}
\text{dim}_{\mathbb{R}}\left( \mathfrak{coll}\left( \mathbb{O}P^{2}\right)
\right) \cong 78=64+14,
\end{equation}%
which leads to the group $\text{Coll}\left( \mathbb{O}P^{2}\right) $ be a $%
E_{6}$ type Lie group as expected.

The same argument is applied for the polarity preserving group $\text{Iso}%
\left( \mathbb{O}P^{2}\right) $, i.e. collineations that preserve also the
elliptic polarity $\pi ^{+}$ or equivalently the form $\beta $. This
argument leads to the Lie algebra $\mathfrak{iso}\left( \mathbb{O}%
P^{2}\right) $ that is given
\begin{equation}
\mathfrak{iso}\left( \mathbb{O}P^{2}\right) =\mathfrak{g}_{2}\oplus
\mathfrak{sa}_{3}\left( \mathbb{O}\right) ,
\end{equation}%
where we intended $\mathfrak{sa}_{3}\left( \mathbb{O}\right) $ the
anti-Hermitian matrices, i.e. $a_{ij}=-a_{ji}^{\ast }$, of null trace.
Elements of this algebra are of the form
\begin{equation}
A=\left(
\begin{array}{ccc}
a_{1}^{1} & a_{2}^{1} & -\left( a_{3}^{1}\right) ^{\ast } \\
-\left( a_{2}^{1}\right) ^{\ast } & a_{2}^{2} & a_{3}^{2} \\
a_{3}^{1} & -\left( a_{3}^{2}\right) ^{\ast } & a_{3}^{3}%
\end{array}%
\right) ,
\end{equation}%
with $a_{3}^{3}=-\left( a_{1}^{1}+a_{2}^{2}\right) $ and $\text{Re}\left(
a_{1}^{1}\right) =\text{Re}\left( a_{2}^{2}\right) =0$. The dimension count
on the generators of the algebra leads to 3 entries of dimension $8$, $2$ of
dimension $7$ and therefore $\text{dim}_{\mathbb{R}}\mathfrak{sa}_{3}\left(
\mathbb{O}\right) =38$ and therefore
\begin{equation}
\text{dim}_{\mathbb{R}}\mathfrak{iso}\left( \mathbb{O}P^{2}\right) \cong
52=38+14,
\end{equation}%
which points to $\text{Iso}\left( \mathbb{O}P^{2}\right) $ as an $F_{4}$%
-type Lie group as expected. With more efforts, following Yokota \cite%
{Yokota}, we can recover all isometry groups giving rise to complex and real
forms of $F_{4}$ (Table \ref{tab:E6 and F4}).
\begin{table}[tbp]
\begin{centering}
\begin{tabular}{|c|c|}
\hline
\textbf{Plane} & \textbf{Isometry group}\tabularnewline
\hline
\hline
$\mathbb{O}P^{2}\left(\mathbb{C}\right)$ & $F_{4}\left(\mathbb{C}\right)$\tabularnewline
\hline
$\mathbb{O}P^{2}$ & $F_{4\left(-52\right)}$\tabularnewline
\hline
$\mathbb{O}_{s}P^{2}$ & $F_{4\left(4\right)}$\tabularnewline
\hline
$\mathbb{O}H^{2}$ & $F_{4\left(-20\right)}$\tabularnewline
\hline
\end{tabular}\caption{\label{tab:E6 and F4}The isometry group of the octonionic and split-octonionic
projective and hyperbolic planes give rise to complex and real forms
of $F_{4}$. }
\par\end{centering}
\end{table}

The same argument can be applied to the elliptic motion group of the
bioctonionic Rosenfeld plane $\text{Iso}\left(\mathbb{\left(\mathbb{C}\otimes%
\mathbb{O}\right)}P^{2}\right)$. Since $\mathbb{C}$ and $\mathbb{O}$ are
both composition algebras, with $\mathfrak{der}\left(\mathbb{C}\right)\cong0$
and $\mathfrak{der}\left(\mathbb{O}\right)\cong\mathfrak{g}_{2}$
respectively, then the Lie algebra of the group of elliptic motion $%
\mathfrak{iso}\left(\mathbb{\left(\mathbb{C}\otimes\mathbb{O}\right)}%
P^{2}\right)$ is given by the direct sum

\begin{equation}
\mathfrak{iso}\left( \mathbb{\left( \mathbb{C}\otimes \mathbb{O}\right) }%
P^{2}\right) \cong \mathfrak{sa}_{3}\left( \mathbb{C}\otimes \mathbb{O}%
\right) \oplus \mathfrak{g}_{2},  \label{eq:decomposition Vinberg}
\end{equation}%
where $\mathfrak{sa}_{3}\left( \mathbb{\mathbb{C}\otimes \mathbb{O}}\right) $
are the anti-hermitian traceless three by three matrices in the bioctonionic
algebra\footnote{%
Brackets on this algebra are not relevant for our argument but can be
derived e.g. from \cite{Vinberg, Barton}.}, i.e. $a_{ij}=-\overline{a}%
_{ji}^{\ast }$ and $a_{00}+a_{11}+a_{22}=0$. Proceeding with the counting on
the generators of the algebra we obtain
\begin{equation}
\text{dim}_{\mathbb{R}}\left( \mathfrak{iso}\left( \mathbb{\left( \mathbb{C}%
\otimes \mathbb{O}\right) }P^{2}\right) \right) =16\times 3+8\times 2+14=78,
\end{equation}%
that gives the well known link between $\mathfrak{\mathfrak{iso}}\left(
\mathbb{\left( \mathbb{C}\otimes \mathbb{O}\right) }P^{2}\right) $ and the
exceptional Lie group $E_{6}$. In the next section we will define all
generalised bioctonionic projective and hyperbolic planes from their
isometry group given as real forms of $E_{6}$ (see Tab. \ref{tab:E6
bioctonionic}).
\begin{table}[tbp]
\begin{centering}
\begin{tabular}{|c|c|}
\hline
\textbf{Plane} & \textbf{Isometry group}\tabularnewline
\hline
\hline
$\left(\mathbb{C}\otimes\mathbb{O}\right)P^{2}$ & $E_{6\left(-78\right)}$\tabularnewline
\hline
$\left(\mathbb{C}_{s}\otimes\mathbb{O}_{s}\right)P^{2}$ & $E_{6\left(6\right)}$\tabularnewline
\hline
$\left(\mathbb{C}\otimes\mathbb{O}_{s}\right)P^{2}$ & $E_{6\left(2\right)}$\tabularnewline
\hline
$\left(\mathbb{C}_{s}\otimes\mathbb{O}\right)P^{2}$ & $E_{6\left(-26\right)}$\tabularnewline
\hline
$\left(\mathbb{C}\otimes\mathbb{O}\right)H^{2}$ & $E_{6\left(-14\right)}$\tabularnewline
\hline
\end{tabular}\caption{\label{tab:E6 bioctonionic}All real forms of $E_{6}$ arise as isometries
of generalised bioctonionic projective and hyperbolic planes.}
\par\end{centering}
\end{table}

\section{\label{6}Bioctonionic Planes as Symmetric Spaces}

\begin{figure}[tbp]
\centering{}\includegraphics[scale=0.7]{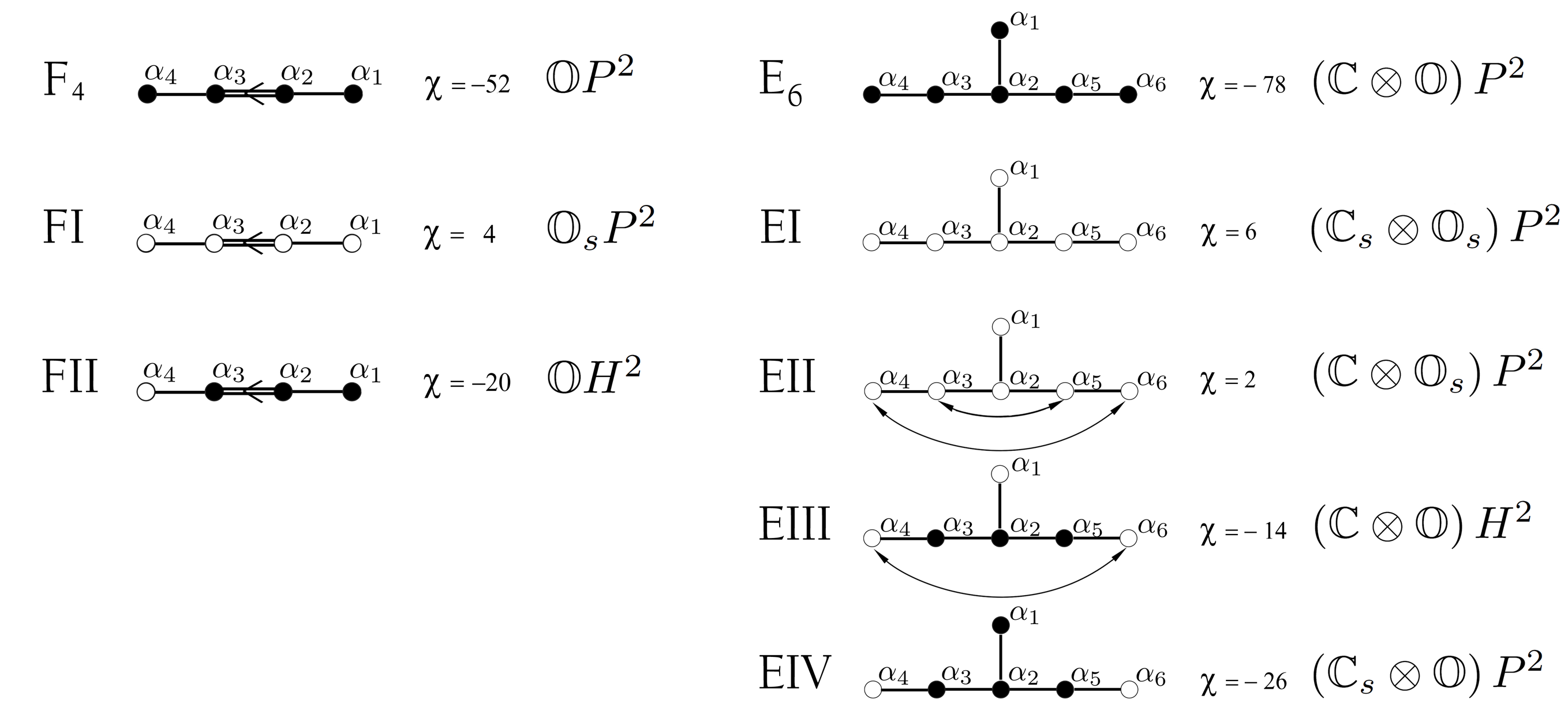}
\caption{Satake diagrams of real forms of $F_{4}$ $E_{6}$, their character $%
\protect\chi$ and corresponding projective plane of which they are the
isometry group. }
\end{figure}
All real forms of rank-3 Magic Squares have been classified and analyzed
e.g. in \cite{key-6} (see also Refs. therein). Moreover, in \cite{magic-pyr}
the $D=3$ layer of the `magic pyramid' of supergravities, containing various
isometry Lie algebras of some Rosenfeld projective planes, is identified
with the $4\times 4$ Lorentzian rank-3 Magic Square $\mathfrak{M}%
_{2,1}\left( \mathbb{A},\mathbb{B}\right) $, where $\mathbb{A}$ and $\mathbb{%
B}$ are the four normed division Hurwitz's algebras \cite{Huruwitz}. It is
here worth remarking that in \cite{key-5} a $6\times 6$ extension of the
Magic Square was also discussed, by introducing null extensions of
quaternions and complex numbers, respectively given by sextonions and
tritonions (see also \cite{Bentz-Dray}).

In the octonionic case, we start from the complexification of the Cayley
plane
\begin{equation}
\mathbb{O}P^{2}\left( \mathbb{C}\right) \simeq \frac{F_{4}(\mathbb{C})}{%
Spin_{9}(\mathbb{C})},
\end{equation}%
and define four different real forms of the plane: one totally compact of
type $(0,16)$ and character $\chi =-16$ identified as $\mathbb{O}P^{2}$ and
that is known as the classical \textit{Cayley plane} or as the \textit{%
octonionic projective plane}; one totally non-compact of type $\left(
16,0\right) $ and character $\chi =16$ identified as $\mathbb{O}H^{2}$ and
known as the \textit{hyperbolic octonionic plane}; and two of type $\left(
8,8\right) $ and character $\chi =0$ named $\mathbb{O}\widetilde{H}^{2}$ and
$\mathbb{O}_{s}\widetilde{H}^{2}$. In all cases the type identifies the
signature, namely the cardinality of non-compact and compact generators,
i.e. $\left( \#_{nc},\#_{c}\right) $, and the character $\chi $ is given by
the difference between the two, i.e. $\chi =\#_{nc}-\#_{c}$. The four plane
are then defined as
\begin{align}
\mathbb{O}P^{2}& \simeq \frac{F_{4(-52)}}{Spin_{9}}, \\
\mathbb{O}H^{2}& \simeq \frac{F_{4(-20)}}{Spin_{9}}, \\
\mathbb{O}\widetilde{H}^{2}& \simeq \frac{F_{4(-20)}}{Spin_{8,1}}, \\
\mathbb{O}_{s}\widetilde{H}^{2}& \simeq \mathbb{O}_{s}P^{2}\simeq \mathbb{O}%
_{s}H^{2}\simeq \frac{F_{4(4)}}{Spin_{5,4}}.
\end{align}

For the bioctonionic case, things are a little more involved and starting
from the complex form of the bioctonionic Rosenfeld plane\footnote{%
The two semispinors $\mathbf{16}_{\mathbb{C}}$ of $Spin(10)_{\mathbb{C}}$ in
the tangent space of (\ref{bioct}) are an example of Jordan pair which is
not made by a pair of Jordan algebras (see e.g. \cite{McCrimmon}).}, i.e.%
\footnote{%
(\ref{bioct}) has a K\"{a}hler structure pertaining to the $\mathbb{C}$
factor in the isotropy/holonomy group.}

\begin{equation}
\left( \mathbb{C}\otimes \mathbb{O}\right) P_{\mathbb{C}}^{2}\simeq \frac{%
E_{6}(\mathbb{C})}{Spin_{10}(\mathbb{C})\otimes \mathbb{C}},  \label{bioct}
\end{equation}%
we have eight real different forms: one totally compact of type $(0,32)$ and
character $\chi =-32$ identified as $\left( \mathbb{C}\otimes \mathbb{O}%
\right) P^{2}$ and that we define as the \textit{bioctonionic Rosenfeld
projective plane}; one totally non-compact of type $(32,0)$ and character $%
\chi =32$ identified as $\left( \mathbb{C}\otimes \mathbb{O}\right) H^{2}$
and that we define as the \textit{bioctonionic Rosenfeld hyperbolic plane};
four plane of type $(16,16)$ and character $\chi =0$ and that are $\left(
\mathbb{C}\otimes \mathbb{O}\right) \widetilde{H}^{2}$, $\left( \mathbb{C}%
\otimes \mathbb{O}_{s}\right) \widetilde{H}^{2}$, $\left( \mathbb{C}%
_{s}\otimes \mathbb{O}\right) \widetilde{H}^{2}$ and $\left( \mathbb{C}%
_{s}\otimes \mathbb{O}_{s}\right) \widetilde{H}^{2}$. The various real forms
of $\left( \mathbb{C}\otimes \mathbb{O}\right) P_{\mathbb{C}}^{2}$ list as
follows \cite{Ros93}:
\begin{align}
\left( \mathbb{C}\otimes \mathbb{O}\right) P^{2}& \simeq \frac{E_{6\left(
-78\right) }}{Spin_{10}\otimes U_{1}}, \\
\left( \mathbb{C}\otimes \mathbb{O}\right) H^{2}& \simeq \frac{E_{6\left(
-14\right) }}{Spin_{10}\otimes U_{1}},  \label{C} \\
\left( \mathbb{C}\otimes \mathbb{O}\right) \widetilde{H}^{2}& \simeq \frac{%
E_{6\left( -14\right) }}{Spin_{8,2}\otimes U_{1}}, \\
\left( \mathbb{C}\otimes \mathbb{O}_{s}\right) P^{2}& \simeq \left( \mathbb{C%
}\otimes \mathbb{O}_{s}\right) H^{2}\simeq \left( \mathbb{C}\otimes \mathbb{O%
}_{s}\right) \widetilde{H}^{2}\simeq \frac{E_{6\left( 2\right) }}{%
Spin_{6,4}\otimes U_{1}}, \\
\left( \mathbb{C}_{s}\otimes \mathbb{O}\right) P^{2}& \simeq \left( \mathbb{C%
}_{s}\otimes \mathbb{O}\right) H^{2}\simeq \left( \mathbb{C}_{s}\otimes
\mathbb{O}\right) \widetilde{H}^{2}\simeq \frac{E_{6\left( -26\right) }}{%
Spin_{5,5}\otimes SO_{1,1}}, \\
\left( \mathbb{C}_{s}\otimes \mathbb{O}_{s}\right) P^{2}& \simeq \left(
\mathbb{C}_{s}\otimes \mathbb{O}_{s}\right) H^{2}\simeq \left( \mathbb{C}%
_{s}\otimes \mathbb{O}_{s}\right) \widetilde{H}^{2}\simeq \frac{E_{6\left(
6\right) }}{Spin_{5,5}\otimes SO_{1,1}}.
\end{align}%
Spaces with $SO_{1,1}$ ($U_1$) factor in the stabilizer are pseudo-K\"{a}%
hler (K\"{a}hler). It is here worth noticing that the Riemannian space $%
\left( \mathbb{C}\otimes \mathbb{O}\right) H^{2}$ (\ref{C}) appears as
enlarged scalar manifold (after 1-form dualization) of $\mathcal{N}=10$, $%
D=2+1$ \textquotedblleft pure\textquotedblright\ supergravity (see e.g. \cite%
{Tollsten}), as well as the non-BPS $Z_{H}=0$ \textquotedblleft moduli
space\textquotedblright\ of extremal black hole attractors in $\mathcal{N}=2$%
, $D=3+1$ exceptional magic theory \cite{Moduli-Spaces}. From the theory of
Jordan triple systems, such a manifold is related to a pair of octonionic
vectors (see \cite{GST2} and Refs. therein).

Finally, there are other two pseudo-Riemannian real forms of the
bioctonionic plane (\ref{bioct}), of type $\left( 20,12\right) $ and $\left(
12,20\right) $, both K\"{a}hler, respectively with character $\chi =8$ and $%
\chi =-8$, namely :%
\begin{eqnarray}
&&\frac{E_{6(2)}}{SO_{10}^{\ast }\otimes U_{1}}, \\
&&\frac{E_{6(-14)}}{SO_{10}^{\ast }\otimes U_{1}},  \label{CC}
\end{eqnarray}%
that apparently do not have a projective or hyperbolic equivalent on
bioctonionic algebras. This fact can be traced back to the absence of the
Lie algebra $\mathfrak{so}_{4n+2}^{\ast }$ in the entries of the real forms
of the Magic Square (cfr. \cite{key-6}, and Refs. therein), and we leave
this intriguing issue for further future work. Here, we only notice that (%
\ref{CC}) appears as the enlarged scalar manifold of $\mathcal{N}=5$, $D=3+1$
\textquotedblleft pure\textquotedblright\ supergravity timelikely reduced to
$\mathcal{N}=10$, $D=3+0$ \textquotedblleft pure\textquotedblright\
supergravity (after complete dualization of 1-forms to 0-forms); cf. \cite%
{BGM}.

\section{\label{7}Musings on the Physics of $E_{6(-78)}$ and $E_{6(2)}$}

The so-called \textquotedblleft exceptional sequence\textquotedblright\ is
given by the Lie algebras $\mathfrak{e}_{n}$ for $n=3,\dots ,8$, which
respectively correspond to $\mathfrak{sl}_{3}\oplus \mathfrak{sl}_{2}$, $%
\mathfrak{sl}_{5}$, $\mathfrak{so}_{10}$, $\mathfrak{e}_{6}$, $\mathfrak{e}%
_{7}$, and $\mathfrak{e}_{8}$. The application of exceptional Lie algebras
in physics was pioneered by G\"{u}rsey. $SU_{5}$ Grand Unified Theories
(GUT) unifies the bosons into a single representation and $Spin_{10}$ GUT
unifies one generation of the fermions, which are contained within $E_{6}$
GUT \cite{Gursey:1975ki}. Bars and G\"{u}naydin explored $E_{8}$ GUT for
three generations of the Standard Model \cite{BarsGunaydin}. While it is
commonly thought that $E_{6}$ is the only exceptional GUT algebra with
complex representations \cite{Witten}, Barr investigated the role of $%
E_{8}\rightarrow E_{6}\otimes SU_{3}$, showing how the $SU_{3}$ flavor
symmetry leads to three generations with mirror fermions and $E_{6}$ GUT
\cite{Barr1988}.

More recently, Dubois-Violette and Torodov explored the state space of three
generations of fermions via $\mathfrak{J}_{3}(\mathbb{O})$ in relation to $%
F_{4}$ \cite{Todorov1}. Boyle elaborated on the role of $E_{6}$ via states
from $\mathfrak{J}_{3}^{\mathbb{C}}(\mathbb{O})$ independent of $E_{6}$ GUT
\cite{BoyleFarnsworth,Boyle}. Krasnov has also discussed the role of $%
\mathbb{O}_{s}\otimes \mathbb{O}$ and $Spin_{11,3}$, but did not obtain
three generations \cite{KrasnovSpin}. Two of the authors have previously
discussed the role of $E_{8(-24)}$ with $Spin_{12,4}$ and a spinor from $(%
\mathbb{O}_{s}\otimes \mathbb{O})\mathbb{P}^{2}$ for three generations of
matter \cite{CMR}.

Wilczek \textit{et al.}~articulate how $Spin_{3}$ flavor symmetry is
preferred over $SU_{3}$ for anomaly cancellation without mirror fermions
\cite{RVW}. $E_{8(-24)}$ contains $Spin_{4,4}\supset Spin_{4,1}\otimes
Spin(3)$, implying extra time dimensions relate to mass eigenstates, as
energy/mass are the time components of energy-momentum in phase space \cite%
{Kostant,CMR}. Wilson found a similar interpretation with $Spin_{3,3}$ and
geometric algebra \cite{Wilson}. Moreover, since $Spin_{6,2}\cong SU_{2,2}(%
\mathbb{H})_{\mathbb{R}}$ (see e.g. \cite{Var}), one can reasonably guess
that $Spin_{4,4}\cong SU_{2,2}(\mathbb{H}_{s})_{\mathbb{R}}$. Thus, it holds
that%
\begin{equation}
\begin{array}{ccc}
Spin_{3,1}\cong SL_{2}(\mathbb{C})_{\mathbb{R}} & \leftarrow &
Spin_{3,3}\cong SL_{2}(\mathbb{H}_{s})_{\mathbb{R}}\cong SL_{4}(\mathbb{R})
\\
\uparrow &  & \uparrow \\
Spin_{4,2}\cong SU_{2,2} & \leftarrow & Spin_{4,4}\cong SU_{2,2}(\mathbb{H}%
_{s})_{\mathbb{R}}\cong Spin(\mathbb{O}_{s})_{\mathbb{R}}%
\end{array}%
\end{equation}%
When singling $SL_{2}(\mathbb{C})_{\mathbb{R}}$ out from $SL_{2}(\mathbb{H}%
_{s})_{\mathbb{R}}$, half of the spinorial degrees of freedom are lost, as $%
\mathbb{H}_{s}$ is made \textit{\`{a} la Cayley-Dickson} from two $\mathbb{C}%
_{s}$ and only contains one $\mathbb{C}$. The triality of $Spin_{4,4}$ leads
to three charts of $Spin_{4,2}\otimes U_{1}$, allowing for three generations
for the price of two in a manner that avoids mirror fermions due to only
containing a single complex representation in $\mathbb{H}_{s}$ via $SL_{2}(%
\mathbb{C})_{\mathbb{R}}\subset SL_{2}(\mathbb{H}_{s})_{\mathbb{R}}$ or $%
SU_{2,2}\subset SU_{2,2}(\mathbb{H}_{s})_{\mathbb{R}}$. A unification of
classical phase space with spacetime and energy-momentum for one generation
can be found in $Spin_{6,2}$ or $Spin_{4,4}$. Therefore, three mass
generations leads to two more energy dimensions, resulting in $Spin_{6,4}$,
a subalgebra of $E_{6(2)}$. The Peirce decomposition of $\mathbf{27}$ w.r.t.~%
$E_{6(2)}$ is used to identify only $\mathbf{16}$ representations as
fermionic, while $E_{6(2)}$ itself also contains spinors; in this framework,
$Spin_{4,4}$ triality can be proposed to describe three generations of
matter.

Recent works discussing $\mathfrak{J}_{3}(\mathbb{O})$ and $\mathfrak{J}%
_{3}^{\mathbb{C}}(\mathbb{O})$ motivate three generations within a single $%
\mathbf{26}$ or $\mathbf{27}$ representation of $F_{4}$ or $E_{6}$ \cite%
{Todorov1,Boyle}. While three charts of $Spin_{9}$ and $Spin_{10}\otimes
U_{1}$ are in $F_{4}$ and $E_{6}$, an appropriate counting of on-shell and
off-shell states must be found. For instance, a complex spinor $\mathbf{16}$
of $Spin_{10}$ is contained within the Peirce decomposition of $\mathbf{27}$
of $E_{6}$. However, three sets of $\mathbf{16}$ as two-component spinors
overlap significantly within $\mathbf{27}$. In other words, we suggest that
the triality of $Spin_{4,4}$, rather than $Spin_{8}$, gives three
generations. $\mathfrak{J}_{3}^{\mathbb{C}}(\mathbb{O})$ cannot fully
contain the field content of three generations of spinors, but Weyl spinors
of $Spin_{4,4}\otimes Spin_{8}$ can, which stems from a single Majorana-Weyl
spinor of $Spin_{12,4}$ within $E_{8(-24)}$ \cite{CMR}.

The three generations of Standard Model fermions within $(\mathbb{O}%
_{s}\otimes \mathbb{O})\mathbb{P}^{2}$ allows for three $(\mathbb{C}\otimes
\mathbb{O})\mathbb{P}^{2}$ spinors, since there are three complex units in $%
\mathbb{O}_{s}$. A single $(\mathbb{C}\otimes \mathbb{O})\mathbb{P}^{2}$
cannot encode three generations, as three combinations of $\mathbf{16}$ are
needed within the theory. If $\mathbf{27}$ contains a mix of fermions and
bosons, rather than $\mathbf{27}$ fermions as found in $E_{6}$ GUT, the $%
\mathbf{27}$ of $E_{6(-78)}$, then the $\mathbf{27}$ as weight vectors does
not have the fermionic roots determined. Once one of three $Spin_{10}\otimes
U_{1}$ is chosen, then, this determines which roots are fermions. A key
departure from $E_{6}$ GUT is the interpretation of the $\mathbf{27}$
representation as purely fermionic; since the fermions of the standard model
are assigned to the pseudo-Riemannian space $E_{8(-24)}/Spin_{12,4}$, the
Peirce decomposition of $\mathbf{27}=\mathbf{16}\oplus \mathbf{10}\oplus
\mathbf{1}$ identifies only the $\mathbf{16}$ as containing fermions. On the
other hand, since $Spin_{10}$ is the largest GUT that contains no additional
fermions, the role of $E_{6(-78)}$ is primarily suggested to encode three
charts of flavor eigenstates of $Spin_{10}\otimes U_{1}$. In this manner,
the utility of $E_{6(-78)}$ is similar if not identical to the one suggested
by Boyle \cite{Boyle}. Our distinction is that only a single generation of
fermions comes from a single $\mathbf{16}\in \mathbf{27}$ of $%
Spin_{10}\subset E_{6(-78)}$; instead, three generations of leptons are
found within\footnote{%
Private correspondence with Todorov confirms that the $\mathbf{27}$ does not
contain three generations, but provides the manifold for three generations.}
$\mathfrak{e}_{6(2)}\oplus \mathbf{27}\oplus \overline{\mathbf{27}}$. Since $%
E_{6(2)}\otimes SU_{3}$ contains the $SU_{3,c}$ of the strong force with
color charge, $Spin_{6,4}\otimes U_{1}\subset E_{6(2)}$ can be identified as
a type of \textquotedblleft gravi-weak\textquotedblright\ symmetry \cite%
{NestiPercacci,Das,Alexander}. On the other hand, the electroweak symmetry $%
SU_{2}\otimes U_1$ is the only subsector of $Spin_{10}$ that is a chiral
gauge theory at low energies, which stems back to spinors of spacetime.
Thus, by putting color aside, $E_{6(2)}$ can be studied to encode mass
eigenstates, which provides a new physical motivation for the non-compact
real form(s) of $E_{6}$ and split octonions. Various paths lead from $%
E_{8(-24)}$ to $SU_{2,2}\otimes U_{1}\otimes SU_{3}\otimes U_{1}$ at low
energies, many of which pass through $SU_{2,2}\otimes U_{1}\otimes
SU_{3}\otimes SU_{2}\otimes U_{1}$ \cite{CMR}.

\section{\label{8}Conclusion}

Making use of the Veronese coordinates, we have explicitly constructed two
different bioctonionic planes, namely the complexification of the Cayley
plane and the bioctonionic Rosenfeld plane, showing some of their different
geometrical features, which yield to different algebraic structures on the
three by three bioctonionic Hermitian matrices. We have also discussed the
isometry groups of the two planes, and then characterized systematically all
possible octonionic and bioctonionic planes as symmetric spaces (of K\"{a}%
hler or pseudo-K\"{a}hler type in the bioctonionic case). This approach
brought us to single out two pseudo-Riemannian real forms of the
bioctonionic plane that apparently do not have a projective or hyperbolic
equivalent on the bioctonionic algebra. One of these spaces appears as
enlarged scalar manifold of a \textquotedblleft pure\textquotedblright\
(i.e. not matter-coupled) supergravity theory with 20 supersymmetries when
dimensionally reduced from $3+1$ to $3+0$ space-time dimensions (namely,
when reduced along time); we plan to investigate more on this in the future.
Finally, we have briefly commented on the physical relevance of some
bioctonionic Rosenfeld plane, hinting to some interesting applications of
real forms of $E_{6}$.

\section*{Acknowledgments}

The work of D.Corradetti is supported by a grant of the
Quantum Gravity Research Institute. The work of AM is supported by a
\textquotedblleft Maria Zambrano" distinguished researcher fellowship,
financed by the European Union within the NextGenerationEU program.

\end{document}